# Resolving observable features of galaxy cluster 'El Gordo' using idealized cluster models in SPH simulations

Erik Husby

Minnesota Institute for Astrophysics, University of Minnesota, 2017

ABSTRACT

The merger scenario of two galaxy subclusters to form the massive galaxy cluster 'El Gordo' is investigated using smoothed particle hydrodynamics (SPH) simulations. Idealized cluster models are used to initialize the states of both subclusters prior to the merger, assuming the commonly used Navarro-Frenk-White (NFW) dark matter (DM) density profile and solving the hydrostatic equilibrium equation for a $\beta$-model with $\beta = 0.6$ to obtain the gas density profile. The impact parameter ($P$) and zero-energy orbit fraction ($\varepsilon$) of the two merging subclusters are varied to put constraints on the space of parameter values which result in projections of X-ray luminosity, Sunyaev-Zel'dovich (SZ) effect, and DM density that correlate well with observational features of 'El Gordo'. We are able to reproduce the remarkable wake-like feature seen in X-ray observations that trails after the secondary (bullet) subcluster as well as the rough shape of the combined cluster with $P \sim 800$ kpc and $\varepsilon = 0.6$, which corroborate the results of prior simulations that support an off-axis, high-speed collision as the merger scenario. We argue that the large separation distance between the mass centers of the two subclusters in this merger scenario may be resolved by finding a better fit for other parameters in the idealized subcluster models.

## INTRODUCTION

The galaxy cluster ACT-CL J0102-4915 currently holds the title of the most massive cluster observed in the distant universe with an estimated mass of $\sim$2-3 $\times$ $10^{15}$ M$_\odot$ (Zhang et al. 2015, Z15 hereafter). Discovered through its strong SZ (thermal) signal by researchers using the Atacama Cosmology Telescope (ACT), a six-meter telescope stationed on Cerro Toco in the Atacama Desert in northern Chile, the cluster was nicknamed 'El Gordo' (meaning "The Fat/Big One" in Spanish) "to reflect its exceptional mass and to recognize the Chilean contribution to ACT" (Menanteau et al. 2012). Further observations of X-ray emission from the cluster using the Chandra telescope revealed a wake-like feature likely created during the passing of a smaller (bullet) cluster through the primary cluster. The large distance between the dark matter cores of the merging subclusters ($\sim$700-800 kpc, approximated by observed offsets between the SZ and X-ray centroids) suggests that 'El Gordo' is the result of a major merger with a high relative velocity of around 2000 km s$^{-1}$. The probability of a massive major merger with such a high relative velocity is extremely low in the standard $\Lambda$CDM cosmological model, making this cluster one of great interest for merger simulations.

Attempts to pin down the most likely merger scenario for the subclusters to reproduce observations of 'El Gordo' have been made through simulations of binary cluster mergers. Setting a small impact parameter $P$ and low zero-energy orbit fraction $\varepsilon$ for the merger scenario (corresponding to a low relative velocity) has reproduced reasonable dark matter core distances between the subclusters with morphology and scaling relations in X-ray and SZ projections that resemble observations using only SPH simulations (Donnert 2014, D14 hereafter). Conversely, setting a large $P$ and relatively high $\varepsilon$ have reproduced a wake-like feature in X-ray projections in addition to reproducing the general morphology and scaling relations using a combination of SPH and the FLASH grid code, with which some cluster substructures such as shocks and eddies can be more accurately simulated (Z15).

We use the best-fit parameter set from Z15 as a base from which we vary $P$ and $\varepsilon$ to test an updated cluster model in simulations of 'El Gordo' using only SPH. Throughout the paper, we assume a concordance cosmology with $H_0 = 70$ km s$^{-1}$ Mpc$^{-1}$, $\Omega_{m,0} = 0.3$, $\Omega_\Lambda = 0.7$ and $\sigma_8 = 0.9$, implying a critical density of $\rho_{crit} = 9.2 \times 10^{-30}$ g cm$^{-3}$ at $z = 0$.

## CLUSTER MODEL

Simulations of galaxy clusters are usually done as part of larger cosmological simulations that take into account resolved substructure in dark matter distribution and its infall into cluster. While this approach offers the most realistic results, the high computational cost of such simulations makes it difficult to explore the parameter space of cluster mergers for analysis. In contrast, direct simulations of merging dark matter haloes initialized with idealized structure are computationally inexpensive and allow for easier analysis of different merger scenarios.

We adopt the approach taken by Donnert et al. (2017) (D17 hereafter) for modeling idealized clusters, using an NFW profile for DM radial density $\rho_{DM}(r)$ and a $\beta$-model for radial gas density $\rho_{gas}(r)$ that are modified to converge at large radii:

$$\rho_{DM}(r) = \frac{\rho_{0,DM}}{\frac{r}{r_s}\left(1 + \frac{r}{r_s}\right)}\left(1 + \frac{r^3}{r_{sample}^3}\right)^{-1} \tag{1}$$

$$\rho_{gas}(r) = \rho_{0,gas}\left(1 + \frac{r^2}{r_c^2}\right)^{-\frac{3}{2}\beta}\left(1 + \frac{r^3}{r_{cut}^3}\right)^{-1} \tag{2}$$

where $r_c$ is the core radius, $\rho_{0,DM}$ and $\rho_{0,gas}$ are the central DM and gas density, respectively, and the NFW scale radius $r_s = r_{200}/c_{200}$. Here, $r_{200}$ is the radius that encloses a mass $M_{200}$ so that the average density $\rho(\langle r_{200}\rangle) = 200\,\rho_{crit}$. Assuming a full halo, the concentration parameter $c_{200}$ varies with cluster mass as (Duffy et al. 2008):

$$c_{200} = 5.74\left(\frac{M_{200}}{2 \times 10^{12}h^{-1}M_\odot}\right)^{-0.097} \tag{3}$$

We model a disturbed cluster assuming (D14) $r_c = \frac{1}{3} r_s$, and set $\beta = 0.6$. The sampling radius for the NFW profile is set to half the simulation box size and $r_{cut} = 1.7\,r_{200}$ (D17). Assuming the canonical baryon fraction ($b_f$) of 17 percent in $r_{200}$, the central DM and gas densities are determined by then assuming the cluster to be in hydrostatic equilibrium.

We model cumulative mass profiles and a temperature profile as (D17):

$$M(<r) = 4\pi \int_0^r \rho(t)t^2 dt \tag{4}$$

$$T(r) = \frac{\mu m_p}{k_B}\frac{G}{\rho_{gas}(r)}\int_r^{R_{max}}\frac{\rho_{gas}(t)}{t^2}M_{tot}(<t)dt \tag{5}$$

## IMPLEMENTATION

We build the idealized cluster models using the latest version of the C code mentioned in D14. The total number of particles used to simulate each cluster is split equally between SPH (gas) and DM particles. Rejection sampling from the total gravitational potential of the gas and DM is performed by interpolating the potential with a cubic spline as done in D17, which allows us to set the initial velocities of DM particles. The initial velocities of gas particles is set to zero.

A low-viscosity simulation scheme is implemented using the latest version of the MHD-SPH code GADGET-3 (Dolag & Stasyszyn 2009; Beck et al. 2016). All models are pre-relaxed in a periodic box for a couple of Gyr to reduce unwanted viscosity caused by density fluctuations in the initial conditions due to Poisson sampling of the density profiles, as explained in D14.

To test the stability of the idealized cluster model, we evolve for 8 Gyr a single cluster of total mass $M_{200} = 1.5 \times 10^{15}$ M$_\odot$ in a periodic box with a side length of 12.8 Mpc. We evolve all models with 5 million DM and 5 million SPH particles. We record the state of all simulations in "snapshot" files at increments of 25 Myr in simulation time. Radial profiles of the single cluster in density, cumulative mass, and temperature are calculated from snapshots taken at both the beginning and end of the 8 Gyr simulation period.

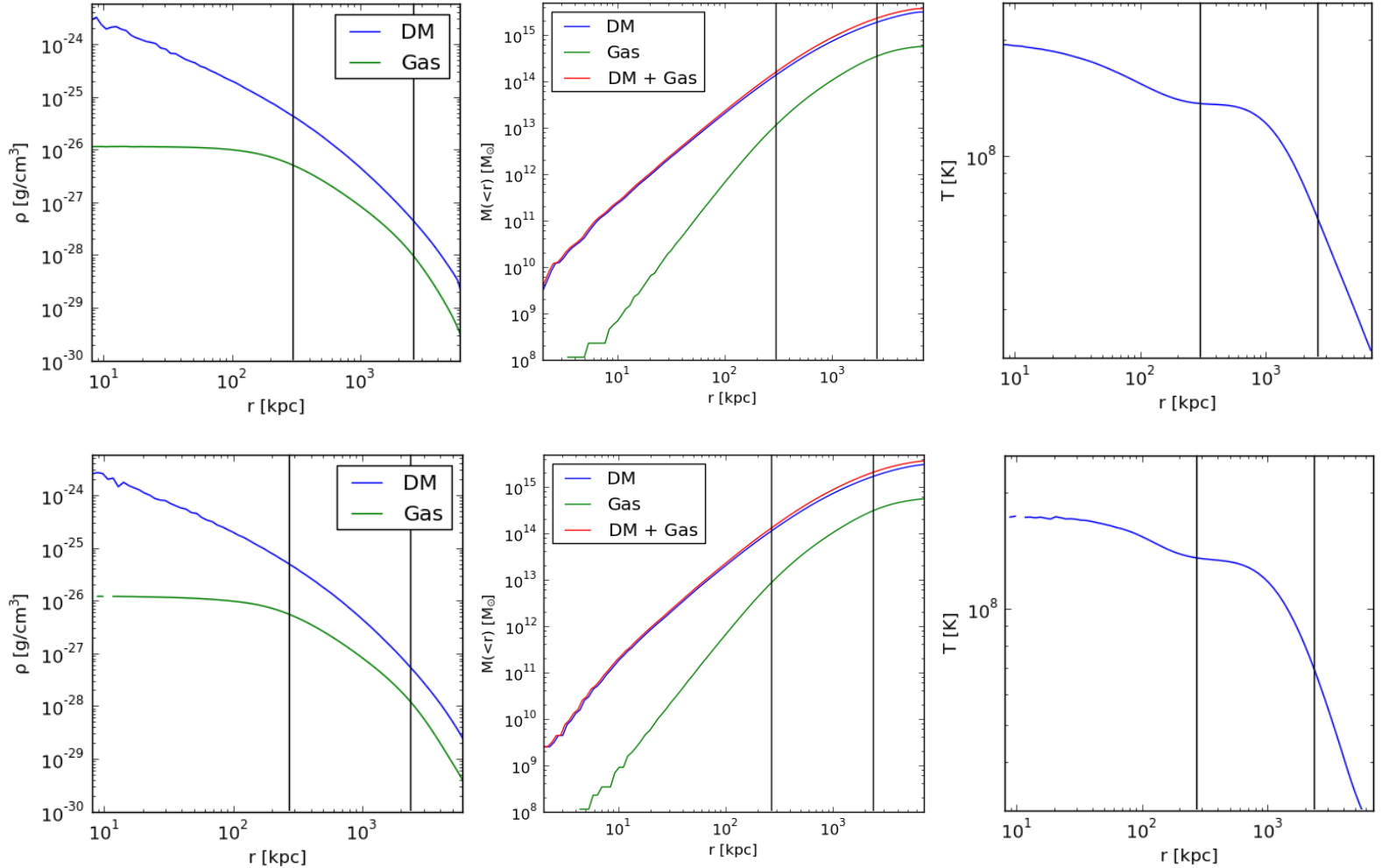

**Figure 1**. Radial profiles for a disturbed cluster with $M_{200} = 1.5 \times 10^{15}$ M$_\odot$. From left to right: DM and gas density, cumulative mass, and temperature. The first row of profiles is from the recorded state of the cluster at the beginning of simulation ($t = 0$), the second row is from the end of the simulation ($t = 8$ Gyr). In each plot, the left and right vertical lines mark the current radial position of $r_c$ and $r_{200}$.

In Fig. 1, we find a good fit of the resulting profiles to those that can be calculated from the analytical models for a cluster of similar $M_{200}$ (see Fig. 1 and 2 in D14, Fig. 3 in D17). The cluster shows stability over the 8 Gyr simulation, with only a slight decrease in radial position of $r_c$ and $r_{200}$ as well as an overall downward shift in the temperature distribution.

## RESULTS

We use the best-fit parameter set from Z15 as a base from which we vary $P$ and $\varepsilon$, holding constant $(M_1, M_2) = (2.5, 0.7) \times 10^{15}$ M$_\odot$ and $(b_{f,1}, b_{f,2}) = (0.05, 0.10)$. Here, $M_1$ is $M_{200}$ of the primary subcluster, $M_2$ is $M_{200}$ of the secondary (bullet) subcluster and $b_{f,1}$, $b_{f,2}$ are the baryon fractions of the two clusters, respectively. It must be noted that our definitions of $M_1$ and $M_2$ differ from those of Z15, where these values correspond to the total mass of the subclusters, due to a clerical error. This results in an overestimation of the total mass of the cluster by a factor of 2, which we assume does not significantly impact the results and the analysis thereof.

To explore the parameter space of $P$ and $\varepsilon$, we simulate the nine different combinations of a low, medium, and high value for each parameter, $\varepsilon \in (0.1, 0.6, 0.9)$ and $P \in (200, 400, 800)$ kpc listed in Table 1. We build the initial conditions of the idealized subclusters in a periodic box with a side length of 15.2 Mpc, then evolve each of the nine variants for 4 Gyr using 256 cores on the Itasca cluster of the Minnesota Supercomputing Institute at the University of Minnesota. All simulations have the merger occur on the xy-plane in coordinate space. We use the C code Smac2 (Donnert & Brunetti 2014) to compute projections of X-ray emission in the 0.5-2.0 keV band to compare with Chandra observations for every snapshot of each merger simulation. These projections are orthogonal to the xy-plane, assuming the merger event of 'El Gordo' to take place in a plane close to the sky plane (Menanteau et al. 2012; Z15). Through visual analysis, we select one snapshot from each of the nine scenarios that best reproduces the X-ray morphology seen in observations as the "best fit" and display them in Fig. 2 in 3x3 matrix form with $\varepsilon$ increasing down rows and $P$ increasing across columns.

We determine the overall best-fit merger scenario of the nine variants to be Variant 6 at time $t = 1.35$ Gyr (V6 hereafter, see the Discussion section), then further explore the parameter space of 3D Euler angles that affect the projection of the snapshot to find a projection for which the subcluster mass center separation is as close as possible to the observed ~700-800 kpc. A 5x5x5 matrix of variants in Euler angles ($E_0$, $E_1$, $E_2$) is used to analyze projected DM surface density and X-ray emission of V6 for reasonable inclination angles of the merger plane away from the sky plane. We construct the matrix such that every combination of the three angles $E_i \in$ (0, 15, 30, 45, 60)°, for $i = 0, 1, 2$ is analyzed and the set of Euler angles that best preserve X-ray morphology resembling observations while bringing the projected mass centers close together is selected. To give the reader an idea of the rotational space that is explored, four X-ray emission projections at the boundaries of the rotational space for Euler angles $E_0$ and $E_1$ are shown in Fig. 3.

**Table 1**. Merger scenario parameter variants for simulations of the El Gordo cluster and the simulation time $t$ of the best-fit snapshot for each scenario. Total X-ray luminosity values $L_{tot}$ calculated by integrating over X-ray projections match the observed $L_{tot} = 2.2 \times 10^{44}$ erg s$^{-1}$ (Menanteau et al. 2012).

| Variant | Parameter | | Best-fit Snapshot | |
|---|---|---|---|---|
| | $\varepsilon$ | $P$ (kpc) | $t$ (Gyr) | $L_{tot}$ ($10^{44}$ erg s$^{-1}$) |
| 1 | 0.1 | 200 | 1.925 | 2.6 |
| 2 | 0.1 | 400 | 1.900 | 2.7 |
| 3 | 0.1 | 800 | 1.900 | 2.8 |
| 4 | 0.6 | 200 | 1.300 | 2.4 |
| 5 | 0.6 | 400 | 1.300 | 2.4 |
| 6 | 0.6 | 800 | 1.350 | 2.4 |
| 7 | 0.9 | 200 | 1.100 | 2.1 |
| 8 | 0.9 | 400 | 1.100 | 2.2 |
| 9 | 0.9 | 800 | 1.125 | 2.2 |

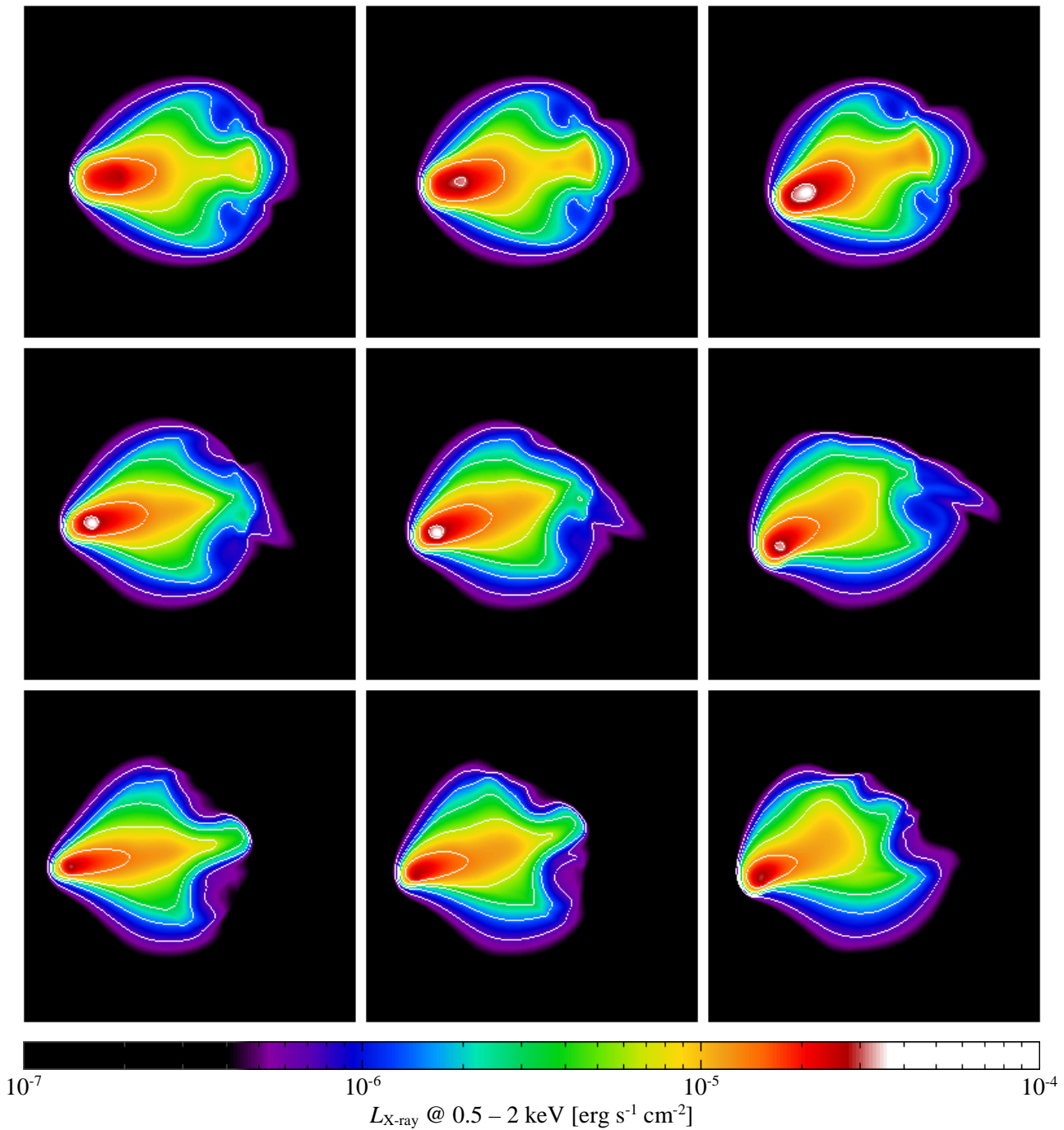

**Figure 2**. X-ray emission projections with 4 Mpc side length of the simulated merging subclusters integrated along a line of sight orthogonal to the merger plane. The white contours range from $1 \times 10^{-6}$ to $32 \times 10^{-6}$ erg/cm$^2$, where the ratio between two successive contours is 2. A single snapshot taken from the 4 Gyr total simulation time for each variant of the model scenario that best replicates observed X-ray features of 'El Gordo' is shown here. The Variant number associated with each model scenario increases from 1 to 9 from top left to bottom right, or rows of this 3x3 parameter variant matrix correspond from top to bottom as $\varepsilon = (0.1, 0.6, 0.9)$ and columns correspond from left to right as $P = (200, 400, 800)$ kpc. Variant 6 in the lower right corner of the figure is selected as the overall best-fit merger scenario of the nine variants (see the Discussion section).

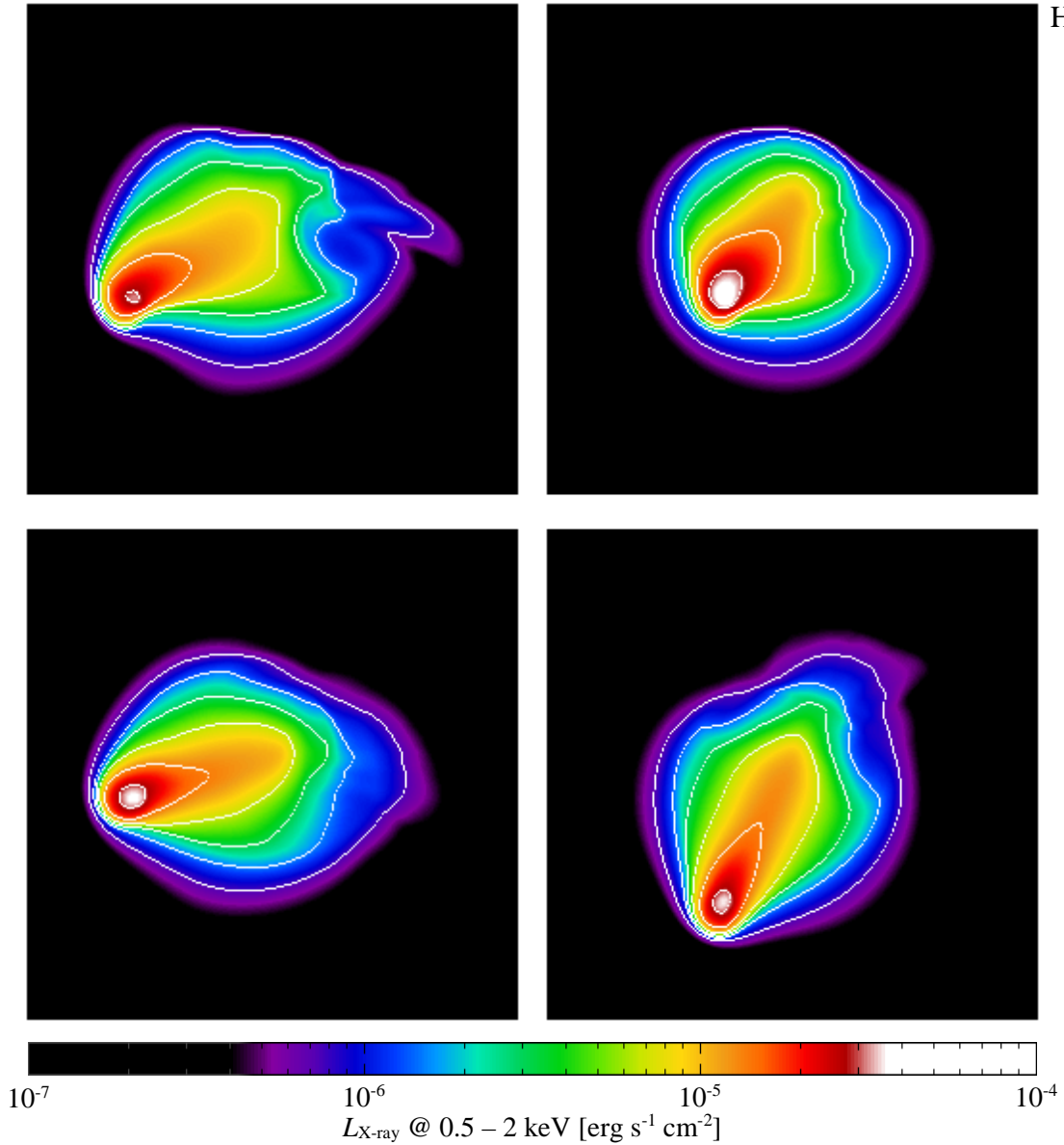

**Figure 3**. X-ray emission projections with 4 Mpc side length using different rotational parameters to change the simulated viewing direction of the best-fit snapshot for Variant 6. The colorbar and contours match those of Fig. 2. The values of Euler angles $(E_0, E_1, E_2)$ are, from top left to bottom right: $(0, 0, 0)°$, $(0, 60, 0)°$, $(60, 0, 0)°$, $(60, 60, 0)°$. The coordinate system used for all merger scenarios has the merger take place in the xy-plane and centered at the origin. Taking the x-axis to be horizontal and the y-axis to be vertical in these projections, for other angles held constant at zero, the effect of increasing one of $(E_0, E_1, E_2)$ is a rotation about the (x, y, z)-axis, respectively. The effect of increasing $E_2$ from the rotational space of bottom right projection may be thought of as a rotation about an axis close to the y = -x axis.

We determine the best-fit set of the 125 different explored combinations of Euler angles for V6 to be $(45, 15, 60)°$ (see the Discussion section) and show projections of X-ray emission, SZ effect, DM surface density, and spectroscopic temperature for this rotation in Fig. 4. A wake-like feature in the projection of X-ray emission remains visible at this viewing direction to the merger, but the separation between the DM cores (mass centers) is about 1280 kpc, considerably larger than the observed value of ~700-800 kpc. The range of the spectroscopic temperature projection is corroborated by the range of temperature values calculated for the single cluster in Fig. 1. The SZ image is smoothed with a Gaussian corresponding to the beam size (FWHM) of ACT at 148 GHz: 1.38 arcmin at z = 0.87 (Swetz et al. 2010).

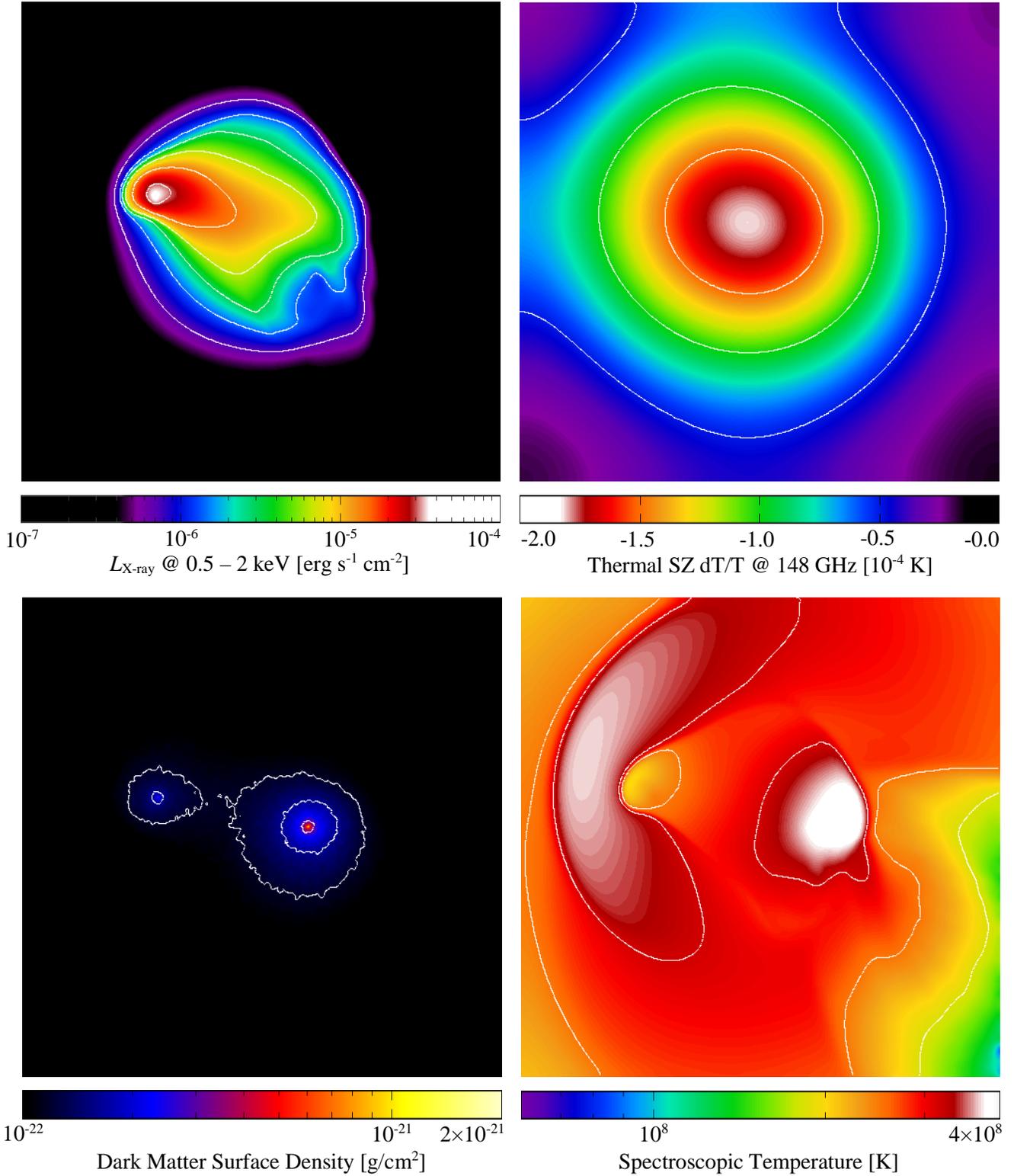

**Figure 4.** Projections of X-ray emission, thermal SZ effect, DM surface density, and spectroscopic temperature (top left to bottom right) of V6 with 4 Mpc side length. The colorbar and contours for X-ray emission are the same as in previous figures, contours for thermal SZ effect are linearly spaced between -2 x $10^{-4}$ K and zero with a separation of $0.5 \times 10^{-4}$ K, the contours for DM surface density range from $2 \times 10^{-21}$ to $3.2 \times 10^{-21}$ g/cm² where the ratio between two successive contours is 2, and contours for spectroscopic temperature range from $6 \times 10^{7}$ K to $36 \times 10^{7}$ K where the amount of increase between the first two contour levels is 2 x $10^{7}$ K and doubles for each subsequent pair of levels.

## DISCUSSION

### Selecting the parameter variant of best fit

To select one snapshot from each of the nine different merger scenarios stemming from parameter variants in impact parameter $P$ and zero-energy orbit fraction $\varepsilon$ that best reproduces the X-ray morphology seen in observations, we scroll through X-ray emission projections (all orthogonal to the merger plane) of each snapshot until the secondary (bullet) subcluster passes through the primary cluster. Following snapshots that show the core of the bullet subcluster just passing through the core of the primary cluster, we plot contours on all projections and search for the next earliest snapshot such that the projection meets the following criteria as best as possible:

(i) The ratios of x-ray luminosity values between contours correlate closely with those between contours on the observations.

(ii) The general positions of contour lines and the spacing between them matches those of observations, within reason.

(iii) A wake-like feature may be seen trailing behind the bullet subcluster.

Using these three criteria, the corresponding X-ray emission projection of the best-fit snapshot for each of the nine scenarios is determined and displayed as in Fig. 2.

We then use the same three criteria to determine the overall best-fit snapshot that most closely resembles observed X-ray morphology of 'El Gordo'. A fourth criterion is added: The general shape of the contour lines is round and features a fair amount of asymmetry as seen in observations. Among the nine projections, those with parameter values $\varepsilon = 0.1$ or $P = 200$ kpc (the first row or first column of Fig. 2, respectively) are too symmetrical. Of the remaining four, Variants 5 and 8 (as listed in Table 1) are still a bit too symmetrical to be the best fit. Of the remaining two, the spacing of contour lines in the outer regions of Variant 9 is too smushed to be the best fit. Thus, Variant 6 (V6) with merger scenario parameters $P = 800$ kpc and $\varepsilon = 0.6$ at a simulation time $t = 1.35$ Gyr is determined to be the best-fit snapshot among those from all tested parameter variants.

### Determining the Euler angles of best rotation

To determine the set of Euler angles $(E_0, E_1, E_2)$ we use to rotate the merger captured in V6 such that the viewing direction best captures the observed features of 'El Gordo', we scroll through projections of both X-ray emission and DM surface mass for all combinations of angles described in the Results section. We search for an X-ray projection that best meets the four criteria mentioned previously in Discussion, with the addition of a fifth criterion: The DM surface mass projection corresponding to the same rotation shows a projected separation of the two subcluster cores that is as close as possible to the approximate observed value ~700-800 kpc. This is done by making X-ray and DM surface mass projections of all explored Euler angle combinations (125 in count) and comparing them side-by-side, checking X-ray projections for proper morphology and DM projections for the distance between the cluster cores.

We find the set of Euler angles $(45, 15, 60)°$ to produce an X-ray emission projection that best meets the criteria for good resemblance to the observed 'El Gordo', including a wake-like structure trailing the bullet subcluster, while attempting to minimize cluster core separation as well. The projected distance between the two cores is about 1280 kpc at this viewing direction, which is still considerably larger than the approximate observed value.

CONCLUSION

We explored the parameter space for modeling the massive binary galaxy cluster merger 'El Gordo'. We presented the NFW profile and $\beta$-profile models that are used to calculate the initial state of an idealized galaxy cluster for its dark matter and gas components, respectively. We showed the stability of the idealized cluster model as it has been implemented to run in SPH simulations by extracting the radial profiles of a single, non-moving cluster and comparing the results to our model.

We ran a series of simulations that evolved different merger scenarios for 'El Gordo', varying the impact parameter and zero-energy orbit fraction to find the set of parameter values that best reproduces the observed features of the cluster. We found that a large impact parameter and a high zero-energy orbit fraction are necessary for our model to reproduce the observed wake-like feature in simulated X-ray emissions, supporting the results of prior studies.

We explored the possibility of the merger plane being at an inclination away from the sky plane and considered the effects this has on the projection of X-ray emission morphology and the projected separation of the two mass centers of the merging subclusters. We find that a separation distance larger than the observed value is necessary to reproduce morphology that resembles 'El Gordo', and acknowledge that more parameters must be varied (such as the concentration parameter of dark matter distribution, the value of $\beta$ in the $\beta$-profile for the gas distribution, and the baryon fraction) in our idealized cluster model to put better constraints on the details of the merger scenario.

ACKOWLEDGEMENTS

The author would like to thank research advisor Julius Donnert for providing valuable insights into the nature of astronomical research and rekindling a love for astronomy in the process.
The author would also like to thank Dinda Khairunnisa for being a pillar of support on this journey.